# Unraveling Privacy Threat Modeling Complexity: Conceptual Privacy Analysis Layers

*Kim Wuyts (PwC Belgium)*
*Avi Douglen (Bounce Security)*

**Abstract** - Analyzing privacy threats in software products is an essential part of software development to ensure systems are privacy-respecting; yet it is still a far from trivial activity. While there have been many advancements in the past decade, they tend to focus on describing 'what' the threats are. What isn't entirely clear yet is 'how' to actually find these threats. Privacy is a complex domain. We propose to use four conceptual layers (feature, ecosystem, business context, and environment) to capture this privacy complexity. These layers can be used as a frame to structure and specify the privacy analysis support in a more tangible and actionable way, thereby improving applicability of the analysis process.

# Privacy Threat Modeling: What can go wrong?

Privacy threat modeling. Technical privacy review. Privacy impact assessment. Privacy engineering analysis. Whatever you like to call it, if you want to understand what can go wrong in a system or feature that is being developed, you will need to analyze at least its security and privacy concerns.

How can you do this best? Well, there are several ways to approach this, from very thorough, rigorous heavy frameworks, to lightweight and lean brainstorming style techniques. They all have one thing in common: a structured approach inspired by the conceptual model you're using. Whether this is your own knowledge in your head that you've built up over the years, the reusable knowledge base from a threat modeling framework, or the hybrid approach combining craft and science; privacy threat knowledge is the essence.

We have seen great advancements in the past 15 years in the privacy threat modeling[1] field. The LINDDUN [1] framework has been updated and extended several times. MITRE's PANOPTIC [2] and Comcast's xCOMPASS [3] have joined the scene. There have been privacy-specific extensions to STRIDE and the Elevation of Privilege [4] card deck: STRIPED [5] and Elevation of Privacy [6]. Furthermore, additional threat libraries have entered that scene that cover a broader scope of threats, such as PLOT4AI [7] which captures security, privacy, and ethics threats targeted at AI systems. And this is just to name a few. This is an active area of research with many new and ongoing developments.

So, does that mean that we know how to look for privacy threats? Yes! And, no! The privacy threat modeling approaches and libraries out there have done a great job in describing *what* privacy threats can be. This is clearly fundamental knowledge for threat elicitation. What is still missing is *how* these threats can be found. This statement needs some nuance. There are, of course, already some great pointers available about the places in a system where certain threats are more likely to occur, or what types of data are or aren't susceptible to certain threat types. Still lacking however is actionable, tangible guidance on how to determine existence or absence of particular privacy issues. You need to understand what a threat looks like and how to find it to know whether it does or doesn't exist.

---

[1] Threat modeling is analyzing representations of a system to highlight concerns about security and privacy characteristics. (Threat Modeling Manifesto, 2020. https://www.threatmodelingmanifesto.org/)

This is actually not surprising, as privacy is a collection of complex concepts and requires more information than the mere user story, feature, or product description to properly analyze. There are different contextual layers to uncover in order to really understand the privacy consequences.

# Privacy Analysis: Layers of Complexity

There are several layers of complexity to privacy analysis. One of the reasons why it might have been so difficult up to now to make the privacy threat elicitation more crisp and tangible is that we have been trying to capture all these different threat indicators in one and the same format (because, that's what roughly[2] has been done for security and it worked there). For privacy, these indicators have varying origins and might require different language to specify.

Let's define these different layers, so they can be used as a frame to better specify the, currently still largely lacking, privacy indicators. Each of the contextual layers will help structure and analyze part of the privacy puzzle. Viewing the functionality through each of these lenses will inform our thinking of the privacy impacts; combining the viewpoints provides a cohesive model to reason about.

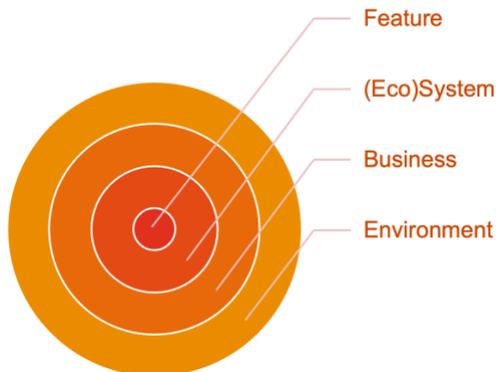

*Fig. 1: The 4 Conceptual Layers to Privacy Threat Modeling*

---

[2] This statement requires some nuance. Security analysis should also take into account business context, etc. A detailed discussion on this would be a paper on its own and is therefore considered out of scope.

## The Feature

*The functionality. The core.*

This is the essence of what we want to analyze. The functionality that is being built. What the system will be doing. This can range from a high-level description of the feature or use case to a more technical model that describes the functionality in more detail (e.g. data flow diagram, sequence diagram, state machine, etc.) This means that we need to understand the user journey (i.e. logical flows) and how it is implemented.

## The (Eco)System

*Putting the feature in the broader context. Interactions with the system as a whole, and the wider technical ecosystem.*

Our feature doesn't exist in isolation. It is part of a bigger system to which it adds additional functionality. This means that it will likely use data that was already part of that existing system in new ways, or will inject new data in the system. It is important that we understand exactly how these interactions work. Understand the bigger picture. And not just focus on the system itself, but also on the external services in this broader ecosystem, e.g. third party products.
This means that we need to understand the underlying system architecture.

## The Business context

*The value. The rationale. The target audience.*

Why are we actually building this feature? What's the value the organization is getting out of this and what's the value for our customers and other stakeholders? We need to understand the purpose of the feature and the individuals involved to assess whether the proposed feature has privacy implications.

This means that we need to understand the business case and its side effects.

## The Environment

*The outside world. External impact on the system.*

The privacy threats and impact will also vary depending on the environment we are in. As an example, location of the target audience will impose legal requirements, cultural expectations, sensitivity of actions or claims, etc. These will all have an effect on the impact of potential privacy threats and therefore need to be taken into account as well.
This means that we need to understand the relevant legislations, their implications, and the individuals' expectations.

## Future Work

Understanding these layers of complexity is a first step. Next, we will need to make each of them more tangible to improve applicability of the privacy analysis process. This will require an interdisciplinary approach to understand the inputs and effects of each of these layers, and a suitable language to properly capture the essence of each layer.

**Inclusion of all contexts in the analysis.** Focusing only on the technical bits and pieces will not be enough for a privacy analysis; you will need to put the feature or product in a broader context. The other way around also applies. While you can get a long way by only focusing on those broader non-technical contextual layers, the threat model gets more value when you also take the more technical details of the system architecture and functionality into account. The non-technical layers may indicate what the threat landscape is, but the technical aspects help determine what threats are impactful to the system.

**Contextual input.** The external layers exceed the typical technical analysis (as known for security threat modeling). We still need to determine how to best capture these indicators and requirements from social sciences, data science, legislation, etc. This will require an interdisciplinary approach.

**Language to represent each layer.** If we want to truly provide support for privacy threat modeling, we will need to revisit what information we need and how we can best express this in an architectural model or other type of representation. Some of these indicators might need a new type of language to capture their essence best.

## Conclusion

Privacy is a complex matter. By understanding its different layers of complexity and leaning into them, we can improve the current state of practice and provide better, more tangible support for structured privacy analysis. This requires a good understanding of the functionality, the broader system context, the business value, and the impact of external factors. Given its multifaceted nature, this will require an interdisciplinary approach where we bring together technical, social, legal, and data science to further evolve this field.

## Acknowledgements

Special thanks to Izar Tarandach and Matthew Coles for challenging, discussing, and improving these ideas.

## Bibliography

[1] DistriNet (KU Leuven), "LINDDUN privacy threat modeling," [Online]. Available: www.linddun.org.

[2] MITRE, "PANOPTIC," [Online]. Available: https://ptmworkshop.gitlab.io/#/panoptic .

[3] Comcast, "xCOMPASS," [Online]. Available: https://github.com/Comcast/xCompass .

[4] A. Shostack, "Elevation of Privilege," [Online]. Available: https://shostack.org/games/elevation-of-privilege .

[5] "STRIPED," [Online]. Available: https://www.youtube.com/watch?v=uzOdpuAhr28 .

[6] "Elevation of Privacy," [Online]. Available: https://github.com/WithSecureOpenSource/elevation-of-privacy .

[7] "PLOT4AI Threat Library," [Online]. Available: https://plot4ai/ .